\documentclass[aps,twocolumn,groupedaddress,showpacs,prb,floatfix]{revtex4}
\usepackage{graphicx,rotating,subfigure,amsmath,amsfonts,amssymb,delarray}
\renewcommand{\vec}[1]{\boldsymbol #1}

\begin{document}
\bibliographystyle{apsrev}


\title{The J1-J2 model: First order phase transition versus deconfinement of spinons}


\author{J. Sirker}
\email[]{sirker@physics.ubc.ca}
\affiliation{Department of Physics and Astronomy, University of British Columbia,
Vancouver, B.C., Canada V6T 1Z1}
\author{Zheng Weihong}
\affiliation{School of Physics, The University of New South Wales,
  Sydney 2052, Australia}
\author{O. P. Sushkov}
\affiliation{School of Physics, The University of New South Wales,
  Sydney 2052, Australia}
\author{J. Oitmaa}
\affiliation{School of Physics, The University of New South Wales,
  Sydney 2052, Australia}

\date{\today}

\begin{abstract}
  We revisit the phase transition from the N\'eel ordered to a valence bond
  solid (VBS) state in the two-dimensional $J_1-J_2$ antiferromagnetic
  Heisenberg model. In the first part we address the question whether or not
  this transition could be an example of a second order phase transition due
  to a deconfinement of spinons. We give arguments based on series expansion
  and spin-wave theory that this is not the case and the transition is most
  likely first order. The method proposed here to detect first order phase
  transitions seems to be very sensitive and might be useful in other models
  as well. In the second part we analyze possible VBS patterns in the
  magnetically disordered phase based on numerical data for different
  susceptibilities, obtained in the ordered phase, which test the breaking of
  lattice symmetries. We conclude that a columnar dimerization pattern is the
  most likely candidate.
\end{abstract}

\maketitle

\section{Introduction}
\label{Intro}
It is well known that microscopically different systems can show similar
behavior near a critical point, a phenomena termed universality. The universal
behavior is caused by the fact that only a small number of long-wave length
degrees of freedom are relevant for most physical quantities at the critical
point. It is indeed often sufficient to consider an effective theory in which
all modes other than the order parameter have been eliminated. The concept of
the order parameter and effective theories based on this quantity has been
developed by Landau and Ginzburg. \cite{LandauLifshitz}

Recently, it has been argued that there is a second order phase transition in
the $S=1/2$ square lattice antiferromagnet between the N\'eel state and a
paramagnetic valence bond solid state which is not described by a
Ginzburg-Landau (GL) type critical theory.
\cite{SenthilViswanath,SenthilBalents} Instead the proposed theory involves
fractional degrees of freedom (spinons) interacting with an emergent gauge
field. One of the best studied models where this scenario could possibly be
realized is the spin-$1/2$ $J_1-J_2$ Heisenberg antiferromagnet in two
dimensions
\begin{equation}
\label{intro1}
H = J_1\sum_{nn} \vec{S}_i\vec{S}_j + J_2\sum_{nnn} \vec{S}_i\vec{S}_j \; .
\end{equation}   
Here $J_1>0$ is the nearest-neighbor interaction and $J_2\geq 0$ a frustrating
next-nearest neighbor exchange. There are two well understood limits: For
$J_1\neq 0, J_2 = 0$ the model is just the usual two-dimensional Heisenberg
antiferromagnet which is known to possess N\'eel order although with an order
parameter $M\approx 0.3$ which is reduced compared to its classical value
$M=1/2$. For $J_1\to 0$ and $J_2\neq 0$ and fixed, on the other hand, both
sub-lattices are N\'eel ordered and $J_1$ then induces a so called collinear
order. For general couplings $J_1,J_2$, both limiting ground states become
frustrated. Therefore a parameter region might be expected where the magnetic
order vanishes and a spin liquid or VBS ground state is formed.  Numerical
studies including exact diagonalization \cite{SchulzZiman}, variational
Quantum-Monte-Carlo \cite{CapriottiSorella,CapriottiBecca} as well as series
expansion
\cite{Gelfand,SinghWeihong,KotovOitmaa,SushkovOitmaa01,SushkovOitmaa02} indeed
indicate that for $0.4\lesssim g\lesssim 0.6$, with $g=J_2/J_1$, no magnetic
order exists.

To address the question of whether or not the ground state in the non-magnetic
region breaks a lattice symmetry, the response to a field
\begin{equation}
\label{intro2}
F_1 = \delta\sum_{i,j} (-1)^i\vec{S}_{i,j}\vec{S}_{i+1,j} 
\end{equation}
has been calculated.
\cite{SushkovOitmaa01,SushkovOitmaa02,CapriottiSorella,CapriottiBeccaED} The
series expansion studies \cite{SushkovOitmaa01,SushkovOitmaa02} show that the
corresponding susceptibility becomes very large or even diverges when
$g\approx 0.4$ is approached from the N\'eel phase, indicating that
translational symmetry by one site is broken and a VBS state is formed. The
results obtained in Ref.~\onlinecite{CapriottiSorella} by a variational quantum
Monte Carlo (QMC) technique seem to support this scenario. Furthermore, the
QMC data have been shown to be in good agreement with an exact diagonalization
of clusters with $N=16, 32$ spins. In a later exact diagonalization study,
however, it has been shown that the susceptibility decreases when going from
the $4\times 4$ to the $6\times 6$-cluster. \cite{CapriottiBeccaED} Further
evidence in favour of a breaking of translational invariance has been obtained
by a dimer series expansion showing directly that the corresponding order
parameter is indeed nonzero in this phase. \cite{KotovOitmaa} On the basis of
the series expansion data we therefore believe that the existence of a
homogeneous spin liquid phase in this parameter region, as proposed in
Ref.~\onlinecite{CapriottiBecca}, is highly unlikely.

A very important result of the series calculations in
Refs.~\onlinecite{SushkovOitmaa01,SushkovOitmaa02,KotovOitmaa} is that the
point $g_{cv}$, where magnetic order vanishes, and the point $g_{c\phi}$,
where dimer order becomes established, are very close or even
coincident. Furthermore, a crossing of energies between an Ising expansion and
a dimer expansion, which would be the indication of a first order transition,
could not be detected. This lead to the assumption that the transition is
second order. For a second order transition it is, however, difficult to
understand within GL theory why $g_{cv}$ and $g_{c\phi}$ should be equal.
Each phase has a different broken symmetry (spin rotational symmetry versus
lattice symmetry) so that one would naively expect that each transition is
described by its own effective theory containing only the staggered
magnetization (order parameter in the N\'eel phase) or the dimer order
parameter, respectively. In this case the transitions should be independent
from each other. That this seems to be not the case tells us that the two
order parameters must be related. A GL-type theory should therefore at least
contain also terms describing the interaction between the two order
parameters.  Such an effective theory has been proposed in
Ref.~\onlinecite{SushkovOitmaa02}. However, it has been found that within this
theory $g_{cv}$ and $g_{c\phi}$ will only be identical if the non-magnetic
phase has massless excitations. This would correspond to a transition to a
translationally invariant spin liquid phase which we believe can be ruled
out based on the numerical data obtained in Ref.~\onlinecite{KotovOitmaa}.

The new effective theory for the second order phase transition between the
N\'eel and a VBS state proposed in
Refs.~\onlinecite{SenthilViswanath,SenthilBalents} follows an entirely
different route. Here the order parameters in the two phases are represented
in terms of fractional degrees of freedom (spinons) which become deconfined
exactly at the critical point. As the spinons are the constituents of both
order parameters this would offer a natural explanation for a direct second
order phase transition between these at first sight very different
phases. Among the models proposed to show such a transition is the
two-dimensional spin-$1/2$ model with a four-spin exchange. \cite{SandvikDaul}
Quite recently, however, it has been argued that the transition in this model
is more likely to be first order. \cite{SpanuBecca} In the present work we
will address the same question about the order of the phase transition for the
$J_1-J_2$ model. 

Our paper is organized as follows: In section \ref{PT} we give arguments based
on series expansion and spin-wave theory why the phase transition is most
likely a weak first order instead of a second order transition. In section
\ref{OP} we will discuss numerical data for three different susceptibilities
probing the VBS order in the non-magnetic phase. These susceptibilities are
obtained in the N\'eel phase where the ground state is known and the series is
therefore not biased. We will discuss why these data provide additional
evidence against the deconfinement scenario and will conclude that the VBS
order is most likely of the columnar dimer type. The last section presents a
summary and conclusions.

\section{Order of the phase transition}
\label{PT}
Usually a first order phase transition is detected in series calculations by
looking for the crossing of energies obtained by expansions starting from
different states. For the $J_1-J_2$ model such an energy-crossing has been
detected between an Ising expansion in the collinear regime and a dimer
expansion in the VBS phase. \cite{KotovOitmaa} This shows that the transition
at $g\sim 0.62$ from the VBS to the collinear state is first order (see also
Fig.~\ref{fig_energies}). For the transition from the N\'eel to the VBS state
at $g\sim 0.4$, on the other hand, no crossing has been found. More precisely,
the energies for an Ising expansion and various dimer expansions are so close
over a relatively large parameter regime around the transition point that it
is not possible to decide within the accuracy of the series if there is a
crossing or not (see Ref.~\onlinecite{KotovOitmaa} and
Fig.~\ref{fig_energies}). In all previous series studies it has been
implicitly assumed that the transition is second order. Here we propose a more
sensitive method to distinguish between first and second order transitions and
conclude that the transition is most likely weak first order.

Let us consider the ground-state energy $e(\delta)$ for the Hamiltonian
(\ref{intro1}) with the field in Eq.~(\ref{intro2}) included for $|\delta|\ll
1$ and $g\leq 0.4$. We have calculated $e(\delta)$ for different $g$ by Ising
series expansion. Using an Ising expansion means that we start with a state
which breaks spin-rotational symmetry whereas the lattice symmetries are
intact. Obviously, it is then impossible to restore spin-rotational symmetry
and break lattice symmetries - as would be required when going from the N\'eel
to the VBS state - in any finite order in the expansion. However, we can
expect that an instability of the state we are starting with is
signaled by a susceptibility, with respect to the corresponding symmetry
breaking field, which is divergent.

\begin{figure}[htbp]
\includegraphics*[width=0.99\columnwidth]{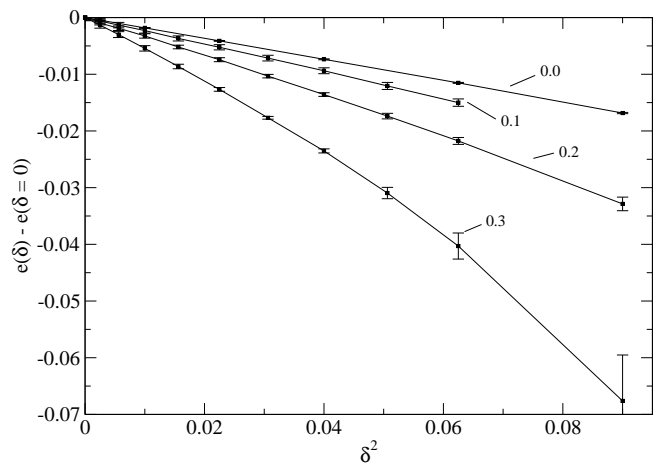}  
\caption{Ising series data for the ground state energy $e(\delta)$ with
  the field $F_1$ included and $g=0.0,0.1,0.2,0.3$. The lines are a guide to
  the eye.}
  \label{fig0}
\end{figure}

\begin{figure}[htbp]
\includegraphics*[width=0.99\columnwidth]{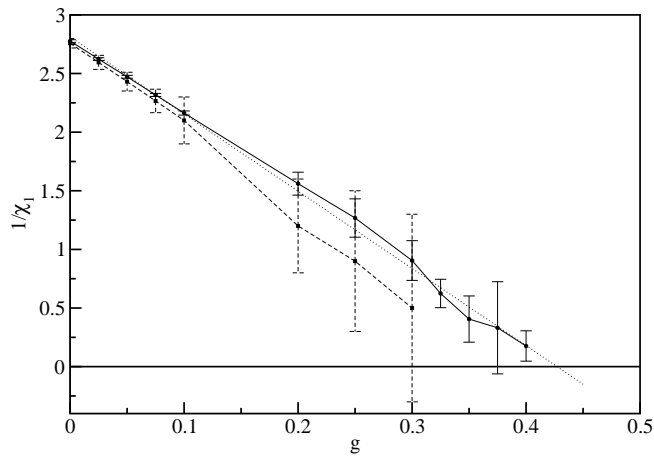}  
\caption{The susceptibility $\chi_1$ calculated as described in the text
  (solid line) compared to the old data from Ref.~\onlinecite{SushkovOitmaa01}
  (dashed line). The dotted line represents a linear fit of the new data
  $1/\chi_1 = 6.58\cdot (0.43-g)$.}
  \label{fig1}
\end{figure}

In Fig.~\ref{fig0} we present our numerical data.
  For a fixed $g$ we have fitted $e(\delta)$ by a polynomial of the form
\begin{equation}
\label{OP_eq1}
e(\delta)-e(0) = \frac{a}{2}\delta^2+\frac{b}{4}\delta^4+\frac{c}{6}\delta^6 \; .
\end{equation}
The susceptibility is then given by
\begin{equation}
\label{OP_eq2}
\chi_1 = -\frac{\partial^2 e}{\partial \delta^2}\bigg|_{\delta=0} = -a \; .
\end{equation}
As the series for the ground state energy shows better convergence than the
series for the susceptibility itself (which was calculated in
Ref.~\onlinecite{SushkovOitmaa01}), we were able to obtain $\chi_1$ with much
smaller error bars than before as shown in Fig.~\ref{fig1}. The new data are
nevertheless consistent with the old data within the given error bars.
A strong response to the field $F_1$ is visible indicating that translational
symmetry is broken in the non-magnetic phase. If the phase transition with
respect to the corresponding order parameter would be second order we expect
$1/\chi_1\propto (g_{c\phi}-g)^{\gamma_\phi}$ where $g_{c\phi}$ is the
critical point and $\gamma_\phi$ the critical exponent. In mean-field theory
$\gamma_\phi = 1$ and we can indeed obtain a nice linear fit of $1/\chi_1$ as
shown in Fig.~\ref{fig1}. Note, that $\gamma_\phi$ is not expected to change
dramatically even if fluctuations are taken into account (an $O(1)$-model, for
example, would have $\gamma_\phi\approx 1.2$) so that changes to the value of
$g_{c\phi}$ would be minor.

To study different possible scenarios for this phase transition we consider an
effective field theory for the magnetically ordered phase. In the effective
field theory for a two-dimensional antiferromagnet in the ordered phase no
topological term (Berry phase) is present. \cite{Haldane2} One can therefore
describe the system by the following $O(3)$-model
\begin{equation}
\label{PT_eq2}
H_{\vec{v}} = \frac{1}{2}\left\{ (\partial_t \vec{v})^2 + c_v^2(\bigtriangledown
  \vec{v})^2+m_v^2\vec{v}^2\right\} +\frac{u_v}{4}(\vec{v}^2)^2 \; .
\end{equation}
Assuming a second order phase transition at a critical point $g_{cv}$ we have
$m_v^2=a_v(g-g_{cv})^{\gamma_v}$ and $u_v>0$. Here $\gamma_v$ is the critical
exponent for the staggered magnetic susceptibility with $\gamma_v = 1$ in
mean-field theory. At $g<g_{cv}$ the vector field $\vec{v}$ will then show a
nonzero ground state expectation value $\langle \vec{v}\rangle
=\sqrt{a_v(g_{cv}-g)/u_v}$.

Consider now the case that we are in the magnetically ordered phase and add
the field $F_1$ as given in Eq.~(\ref{intro2}) with $|\delta|\ll 1$. The
N\'eel order will then coexist with a small dimerization described by a scalar
field
\begin{equation}
\label{PT_eq3}
H_\phi = \frac{1}{2}\left\{ (\partial_t \phi)^2 + c_\phi^2(\bigtriangledown
  \phi)^2+m_\phi^2\phi^2\right\} +\frac{u_\phi}{4}\phi^4+\frac{r_\phi}{6}\phi^6-\delta\phi \; .
\end{equation}
If a second order phase transition with respect to $\phi$ at a critical point
$g_{c\phi}$ would occur, we would have
$m_\phi^2=a_\phi(g_{c\phi}-g)^{\gamma_\phi}$ and $u_\phi>0$. We also want to
include an interaction between the vector and the scalar field. The lowest
order coupling term allowed by symmetry is
\begin{equation}
\label{PT_eq4}
H_{\text{int}} = \frac{u_{v\phi}}{2}\vec{v}^2\phi^2 \; .
\end{equation}
The effective field theory in the ordered phase for $\delta\neq 0$ is then
given by $H=H_v+H_\phi+H_{\text{int}}$ and we will have a nonzero ground state
expectation value 
\begin{equation}
\label{PT_eq5}
\langle\phi\rangle = \frac{\delta}{A}-\frac{u_\phi}{A^4}\delta^3+\frac{3u_\phi^2-Ar_\phi}{A^7}\delta^5+\mathcal{O}(\delta^7) 
\end{equation}
with $A=m_\phi+u_{v\phi}\langle \vec{v}\rangle^2$. This leads to a ground
state energy given by
\begin{equation}
\label{PT_eq6}
e(\delta)-e(\delta=0) = -\frac{1}{2A}\delta^2+\frac{u_\phi}{4A^4}\delta^4+\frac{Ar_\phi-3u_\phi^2}{6A^7}\delta^6+\mathcal{O}(\delta^8) \; .
\end{equation}
This regular expansion in powers of $\delta^2$ will exist for any
$g<g_{c\phi}$ but the radius of convergence will become smaller and smaller
when the assumed critical point is approached. Finally, this expansion will
break down and directly at the critical point $e(\delta)$ will show a scaling
with a critical exponent which is in general non-integer. Nevertheless, in the
parameter regime where this expansion is valid we expect the coefficient of
the $\delta^4$-term to be {\it positive} because $u_\phi>0$ for a second order
phase transition.

\begin{figure}[htbp]
\includegraphics*[width=0.99\columnwidth]{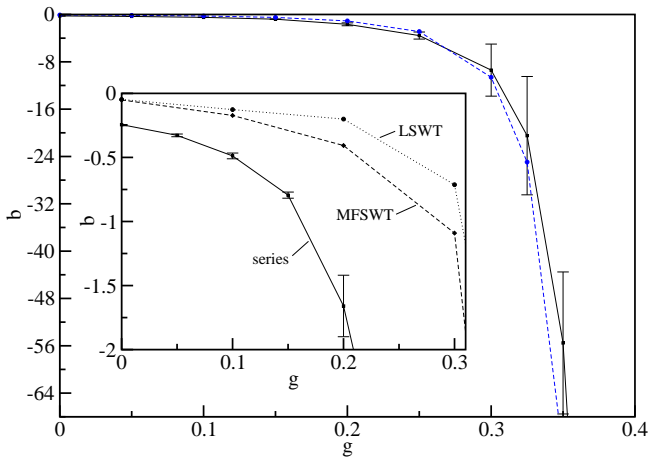}  
\caption{The coefficient $b$ of the quartic term in Eq.~(\ref{OP_eq1})
  obtained from a fit of the data in Fig.~\ref{fig0} (black squares). The
  dashed line represents a fit $b(g) = -5.68/[6.58(g-0.43)]^4$. Inset: The
  coefficient $b$ from LSWT and MFSWT in comparison to the series data. The
  lines are a guide to the eye.}
  \label{fig6}
\end{figure}

We have fitted $e(\delta)$ shown in Fig.~\ref{fig0} by the polynomial given in
Eq.~(\ref{OP_eq1}) but studied so far only the coefficient of the quadratic
term which gives the susceptibility (see Fig.~\ref{fig1}). The coefficient of
the quartic term obtained by a fit of these data is shown in Fig.~\ref{fig6}.
Surprisingly, this coefficient is negative. Before we discuss the consequences
we will check if the coefficient follows the form predicted in (\ref{PT_eq6}).
The coefficient $A=1/\chi_1$ so that we can use the linear fit shown in
Fig.~\ref{fig1} for this parameter. With these values for $A$ the data in
Fig.~\ref{fig6} are in very good agreement with Eq.~(\ref{PT_eq6}) where the
single parameter fit (shown as dashed line) yields $u_\phi=-5.68$.

A check of our series data for $b$ is provided by calculating this coefficient
in spin-wave theory.  We know that linear spin-wave theory (LSWT) yields
accurate results for quantities like ground-state energy or sub-lattice
magnetization for the two-dimensional Heisenberg antiferromagnet without
frustration.\cite{Anderson,Takahashi} Surprisingly, LSWT gives also a phase
diagram for the $J_1-J_2$ model \cite{ChandraDoucot} which is very similar to
the one found by numerical calculations. In particular, a non-magnetic ground
state for $g\in [0.38,0.51]$ is found. In a spin-wave theory where the quartic
terms are treated self-consistently in one-loop approximation (MFSWT),
however, the phase diagram changes dramatically and N\'eel order remains
stable up to $g\sim 0.6$.\cite{XuTing} This tells us that we cannot trust
spin-wave theory in the strongly frustrated regime. If we consider $g$ close
to zero, on the other hand, we might expect that spin-wave theory results are
reasonable.

In complete analogy to the series calculations we have calculated the
ground-state energy $e(\delta)$ in LSWT and MFSWT for the $J_1-J_2$ model with
the term (\ref{intro2}) included. The coefficient of the $\delta^4$-term
obtained from these calculations is shown in the inset of Fig.~\ref{fig6} in
comparison to the series data. Again we find that $b$ is negative and
decreases with increasing $g$.  Quantitatively, $b$ is larger, both in LSWT
and MFSWT, as in series and decreases more slowly.

A negative $u_\phi$ in Eq.~(\ref{PT_eq3}) means that the phase transition with
respect to $\phi$ will be {\it first order} and the assumed critical point
$g_{c\phi}$ will never be reached. For $\delta=0$ the order parameter
$\langle\phi\rangle$ will instead jump from zero to some finite value. To
determine the transition point one also needs to know $r_\phi$ which has to be
positive for stability reasons. In principle, $r_\phi$ can be determined from
the sixth order term in (\ref{PT_eq6}). In practice, the errors in the
numerical data for $e(\delta)$ and in the parameters $A, u_\phi$ are too large
to determine $r_\phi$ reliably. Due to the interaction term (\ref{PT_eq4}) a
jump in $\langle\phi\rangle$ will induce a simultaneous jump in the N\'eel
order parameter $\langle\vec{v}\rangle$. Depending on the strength of the
interaction $u_{v\phi}$ two scenarios are possible: The N\'eel order parameter
could jump to zero yielding a direct first order transition from the ordered
phase to a disordered phase with broken translational invariance. A more
exotic scenario, where $\langle\vec{v}\rangle$ jumps to a smaller but nonzero
value and then decreases further before vanishing at a critical point, is also
possible. This would imply that there is a region in the phase diagram where
dimerization and N\'eel order coexist. Based on the currently available
numerical data it is impossible to decide which scenario is actually realized.
We also want to mention that our mean-field treatment of the order parameters
and the use of the expansion (\ref{PT_eq6}) are a posteriori justified. The
finding of a first order phase transition means that we never get to the
assumed critical point $g_{c\phi}$ where the length scale for fluctuations
would diverge and a mean-field treatment would therefore be very questionable.
It also means that the expansion (\ref{PT_eq6}) exists and its radius of
convergence will be finite everywhere in the ordered phase.

\section{VBS order in the non-magnetic phase}
\label{OP}
In this section we provide additional, independent arguments against the
deconfinement scenario and in favor of a first order phase transition. We also
address the question which kind of VBS order is actually realized.

\begin{figure}[htbp]
\includegraphics*[width=0.99\columnwidth]{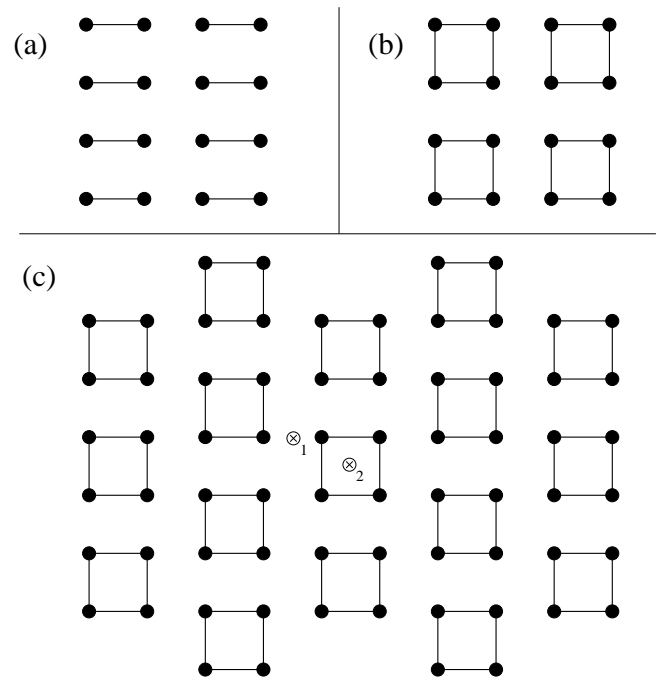}  
\caption{Possible VBS ordering patterns: (a) columnar dimer order, (b) plaquette
  order and (c) plaquette order with every second column of plaquettes shifted
  by one lattice site.}
  \label{fig4}
\end{figure}

In Refs.~\onlinecite{SenthilViswanath,SenthilBalents} dealing with a possible
deconfined critical point separating a N\'eel ordered from a VBS phase in
a two-dimensional antiferromagnet it has been implicitly assumed that the VBS
order is either of columnar dimer type as shown in Fig.~\ref{fig4}(a) or of
plaquette type as shown in Fig.~\ref{fig4}(b).
The order parameter for columnar dimerization is given by $O_c = (-1)^i
\vec{S}_{i,j}\vec{S}_{i+1,j}$ whereas the order parameter for the plaquette
phase can be represented as $O_p = (-1)^i \vec{S}_{i,j}\vec{S}_{i+1,j}+(-1)^j
\vec{S}_{i,j}\vec{S}_{i,j+1}$. As discussed in
Ref.~\onlinecite{SenthilBalents} they can also be interpreted as a single
complex order parameter where only the phase is different for the two patterns.
At a deconfined critical point, however, this phase will only appear as an
irrelevant operator. Therefore {\it both} order parameters are expected to
show power law correlations at such a point and susceptibilities testing the
breaking of lattice symmetries with respect to columnar or plaquette order
should diverge when the deconfined critical point is approached from the
magnetically ordered phase.

\begin{figure}[htbp]
\includegraphics*[width=0.99\columnwidth]{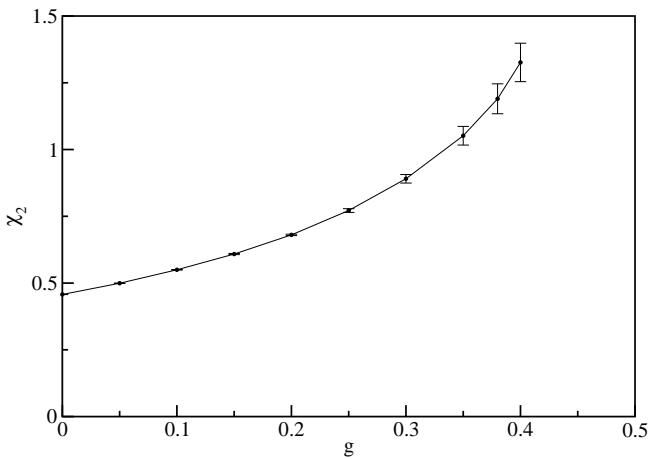}  
\caption{Susceptibility $\chi_2$ as a function of $g$ calculated by Ising
  series expansion. The line is a guide to the eye.}
  \label{fig2}
\end{figure}

First, consider the field
\begin{equation}
\label{OP_eq3}
F_2 = \delta \sum_{i,j} \left(\vec{S}_{i,j}\vec{S}_{i+1,j}-\vec{S}_{i,j}\vec{S}_{i,j+1}\right) 
\end{equation}
which tests if rotational symmetry is broken, which would be the case for the
columnar dimer state but not for the plaquette state. The corresponding
susceptibility $\chi_2$ shows only a very moderate increase with $g$ and
certainly no sign of divergence when the possible critical point
$g_{c\phi}\approx 0.43$ is approached (see Fig.~\ref{fig2}).
We mention here that our results for the susceptibilities $\chi_1$ and
$\chi_2$ are in qualitative agreement with Ref.~\onlinecite{CapriottiSorella}
where the finite size scaling of these susceptibilities has been investigated
by exact diagonalization and variational QMC methods. Note that it does not
come as a surprise that there is no quantitative agreement: the variational
QMC method explores only a small part of the Hilbert space so that a
convergence to the true value for the susceptibility cannot be expected even
if arbitrary large clusters could be treated. If a good variational
wave-function is chosen, however, one might expect that the finite size
scaling is at least qualitatively similar to the true finite size scaling.  

\begin{figure}[htbp]
\includegraphics*[width=0.99\columnwidth]{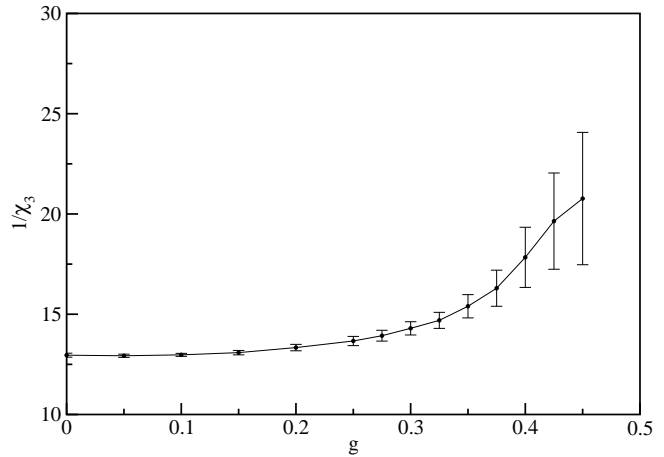}  
\caption{Susceptibility $\chi_3$ as a function of $g$ calculated by Ising
  series expansion. The line is a guide to the eye.}
  \label{fig3}
\end{figure}

Next, we consider the susceptibility $\chi_3$ corresponding to the field
\begin{equation}
\label{OP_eq4}
F_3 = \delta \sum_{i,j} (-1)^{i+j}\left(S^x_{i,j}S^x_{i+1,j+1}+S^y_{i,j}S^y_{i+1,j+1}\right) 
\end{equation}
which has already been calculated in Ref.~\onlinecite{SushkovOitmaa02}. We
reproduce the result in Fig.~\ref{fig3}.
$\chi_3$ even {\it decreases} when the non-magnetic phase is approached which
seems to indicate that the corresponding lattice symmetry, which would be
broken for the plaquette but not for the columnar dimer state, is intact.

The fact that $\chi_2$ and $\chi_3$ do not diverge when $g\rightarrow
g_{c\phi}\approx 0.43$ as would be expected for a deconfined critical point
provides an argument independent of the considerations in section \ref{PT}
that the phase transition in the $J_1-J_2$ model is not an example for the
scenario proposed in Refs.~\onlinecite{SenthilViswanath,SenthilBalents}.
Putting the arguments given in section \ref{PT} aside, this alone, however,
does not exclude the possibility of another kind of second order phase
transition from the N\'eel ordered to a state with VBS order. Nevertheless, in
any second order scenario we should take the fact that $\chi_2$ and $\chi_3$
do not diverge seriously.  This means that in this case the VBS order cannot be
of columnar dimer type because then $\chi_2$ should diverge, and not of
plaquette type because this would lead to a diverging $\chi_3$. So in such a
scenario the quest is to find an ordering pattern with lattice symmetries
which agree with our findings for all three susceptibilities.

A possible pattern is shown in Fig.~\ref{fig4}(c).  Translational symmetry by
one site along the x-axis is broken leading to a divergent $\chi_1$.  $\chi_3$
will be finite because of the $180^o$ rotational symmetry around the axis
marked by ``1'' in Fig.~\ref{fig4}(c) or identical positions. $\chi_2$ will be
finite for this pattern because of the following symmetry: A rotation by
$90^o$ around the axis ``2'' or identical positions, followed by a shift of
every second row of plaquettes by one lattice site, followed by a shift of
every second column of plaquettes by one lattice site.  Although other pattern
which have the correct symmetries might be possible, the pattern in
Fig.~\ref{fig4}(c) seems to be the one with the smallest unit cell.

We have calculated the ground-state energy as well as the singlet and triplet
dispersion for different $g$ for this state by series expansion starting from
the decoupled plaquettes shown in Fig.~\ref{fig4}(c). The expansion parameter
$x$ describes the couplings between the plaquettes. For $x=0$ the plaquettes
are decoupled whereas $x=1$ corresponds to the case we are finally interested
in where the couplings within and between the plaquettes are of equal
strength.

\begin{figure}[htbp]
\includegraphics*[width=0.99\columnwidth]{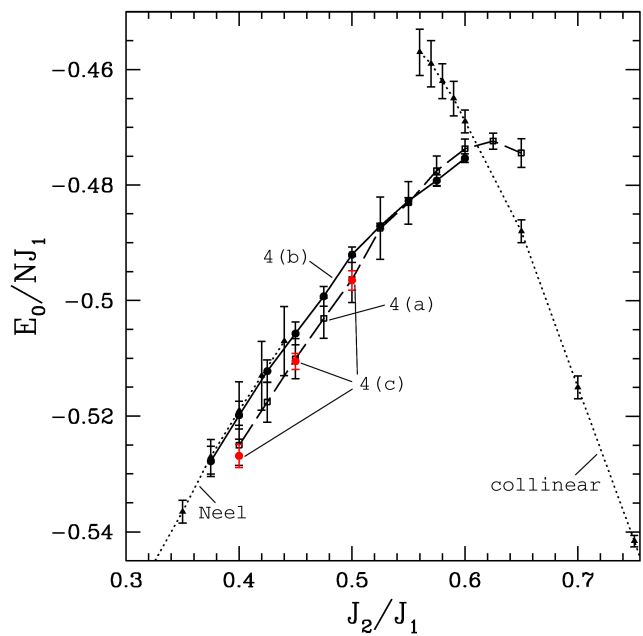}  
\caption{Ground state energies calculated by different series expansions starting
  from the nonmagnetic states shown in Fig.~\ref{fig4} as well as from the
  magnetic N\'eel and collinear states. The curve 4(a) (open squares)
  corresponds to the columnar dimer state shown in Fig.~\ref{fig4}(a), the
  curve 4(b) (black dots) to the plaquette state shown in Fig.~\ref{fig4}(b),
  and the curve 4(c) (red dots) to the plaquette state shown in
  Fig.~\ref{fig4}(c). The black triangles show results of Ising series
  expansions starting from the N\'eel and collinear states.}
\label{fig_energies}
\end{figure}

In Fig.~\ref{fig_energies} we show the ground-state energy obtained from this
expansion compared to other series data.

The obtained energies for $0.4\leq g\leq 0.5$ are similar or even slightly
lower than the energies obtained from the columnar dimer expansion. This shows
that the state in Fig.~\ref{fig4}(c) is indeed a possible candidate for the
ground state in this parameter region. In series calculations it is, however,
important to make sure that the state which is the starting point for the
expansion remains stable when extrapolating to the isotropic limit
$x\rightarrow 1$. We have therefore calculated also the singlet and triplet
dispersions. The plaquettes for this VBS state are arranged on a triangular
lattice. With respect to this lattice both singlet and triplet dispersion show
a mininum at momentum $\vec{k} = (0,2\pi/\sqrt{3})$. The series coefficients
for the singlet and triplet gaps at this point and $g=0.25, 0.45$ are given in
Table \ref{tab_ser}. 
\begin{table*}[htbp]
\squeezetable
\caption{Series coefficients $x^n$ for the minimum singlet gap $\Delta_s$ and
  triplet gap $\Delta_t$ for the plaquette state from Fig.~\ref{fig4}(c) and
  $g=0.25$ ($g=0.45$), respectively.}
\label{tab_ser}
\begin{ruledtabular}
\begin{tabular}{ccccc} 
\multicolumn{1}{c}{$n$} &\multicolumn{1}{c}{$\Delta_s/J_1\; (g=0.25)$} 
&\multicolumn{1}{c}{$\Delta_t/J_1\; (g=0.25)$} &\multicolumn{1}{c}{$\Delta_s/J_1\; (g=0.45)$} 
&\multicolumn{1}{c}{$\Delta_t/J_1\; (g=0.45)$} \\
\hline
0 & 1.500000000 & 1.000000000 & 1.100000000 & 1.000000000\\
1 & 0.000000000 & -8.33333334$\times 10^{-1}$ & 0.000000000 & -4.33333334$\times 10^{-1}$\\
2 & -4.356195887$\times 10^{-1}$ & -5.722808442$\times 10^{-1}$ & -4.151003339$\times 10^{-1}$ & -5.005966374$\times 10^{-1}$\\
3 & -3.880918282$\times 10^{-1}$ & 1.4032850965$\times 10^{-1}$ & -1.832689336$\times 10^{-1}$ & 8.2113320127$\times 10^{-2}$\\
4 & -4.828163654$\times 10^{-1}$ & -2.265506840$\times 10^{-1}$ & -1.537390367$\times 10^{-1}$ & -3.080455443$\times 10^{-2}$\\
5 & -6.032418702$\times 10^{-1}$ & 7.5790435073$\times 10^{-3}$ & -1.592497102$\times 10^{-1}$ & -1.108096117$\times 10^{-1}$\\
6 & -7.988310579$\times 10^{-1}$ & -4.517676083$\times 10^{-2}$ & -1.653261118$\times 10^{-1}$ & 1.6084176410$\times 10^{-2}$\\
\end{tabular}                                                       
\end{ruledtabular}                                                  
\end{table*}
Dlog Pad\'e approximants for these series show that the singlet gap always
vanishes before the triplet gap and at a value $x_c<1$. This indicates that
the state in Fig.~\ref{fig4}(c) is unstable.

For the columnar dimer state, on the other hand, it is very difficult to
obtain the singlet dispersion by series expansion because the state we are
starting with consisting of decoupled dimers does not contain a singlet
excitation. We have, however, calculated the triplet gap in this case which
shows a minimum at $\vec{k}=(0,\pi)$ and Dlog Pad\'e approximants suggest that in
this case the gap remains nonzero for $x\rightarrow 1$. In addition, the
instability of the state in Fig.~\ref{fig4}(c) which manifests itself by a
vanishing of the singlet gap at $\vec{k} = (0,2\pi/\sqrt{3})$ for $x<1$ is an
instability towards a columnar dimerization. Although this does not prove that
the columnar state is finally stable, it makes the columnar dimerization
pattern the most likely candidate for the VBS order in the non-magnetic region
of the phase diagram.

\section{Conclusions}
\label{Conclusion}
We have calculated the ground-state energy $e$ for the $J_1-J_2$ model
including a small field $F_1$ with strength $\delta$, which induces a columnar
dimerization in the N\'eel ordered phase, by Ising series expansion. We have
argued that everywhere except directly at a critical point it is possible to
expand $e(\delta)$ in a regular series in $\delta^2$ and that this series has
a finite radius of convergence which goes to zero when the tentative critical
point is approached. The prefactor of the $\delta^2$-term in that series gives
the susceptibility $\chi_1$ with respect to $F_1$. The data for this
susceptibility obtained by an Ising series expansion indicate that the N\'eel
state becomes unstable for $g>g_c\approx 0.43$ and that the ground state for
$g>g_c$ breaks translational symmetry by one site and therefore seems to be of
the VBS type and not a spin liquid. Based on a mean-field treatment of an
effective field theory describing the N\'eel state coexisting with the small
dimerization induced by $F_1$ we have argued that the sign of the
$\delta^4$-term in the expansion of $e(\delta)$ determines whether the
transition with respect to the VBS order parameter is first or second order.
We believe that this is an in general more sensitive and less biased method to
distinguish between a first and a second order transition than looking for a
crossing of energies obtained by different expansions. From the series data we
found that the $\delta^4$-term has a negative sign and we showed that the same
is true in spin-wave theory. Within the presented GL-type theory this means
that the transition is expected to be first order.  Our mean-field treatment
of the order parameter is a posteriori justified because a critical point
where such a treatment would break down is never reached.

In the second part we gave arguments in favor of a first order transition
which are independent of any effective field theory by analyzing two
additional susceptibilities testing different lattice symmetries. These
susceptibilities were calculated based on an Ising expansion in the N\'eel
phase so that the series is not biased by any assumed dimerization pattern. We
argued that at a deconfined critical point all three susceptibilities
considered here are expected to diverge and that the fact that $\chi_2$ and
$\chi_3$ do not diverge excludes this scenario. We further argued that in any
second order scenario the non-divergence of $\chi_2$ would mean that the VBS
state is not of the columnar dimer type and the non-divergence of $\chi_3$
would mean that the VBS state is not of plaquette type either. For an assumed
second order transition we have been able to find a VBS pattern which does
have the correct lattice symmetries to explain our data for all three
susceptibilities. Series expansion data starting from this pattern, however,
have proven that this state is unstable with respect to the columnar
dimerization pattern.

Taking the arguments given in the two parts together shows that the transition
from the N\'eel state to a VBS state with columnar dimerization is most likely
first order. For a first order phase transition we still expect that the
susceptibility $\chi_1$ when calculated by Ising series expansion diverges
because the field $F_1$ directly tests the instability of the N\'eel state
with respect to columnar dimer order irrespective of the order of the phase
transition. For the susceptibility $\chi_2$, on the other hand, we can only
expect that it diverges if the transition is second order. The fact that
$\chi_2$ does increase with $g$ without diverging at $g\approx 0.4$ therefore
further supports that the transition is most likely first order and the VBS
state most likely of the columnar dimer type.
\begin{acknowledgments}
The authors acknowledge valuable discussions with I.~Affleck. JS acknowledges
support by the German Research Council ({\it Deutsche
  Forschungsgemeinschaft}). 
\end{acknowledgments}

\end{document}